\documentclass[conference]{IEEEtran}

\IEEEoverridecommandlockouts
\usepackage{tabularx,graphicx}
\usepackage{subfigure}
\usepackage{amsmath,amssymb,amsfonts}
\usepackage{amsthm}
\usepackage{mathtools,newtxmath}
\usepackage{algorithm, algorithmicx,algpseudocode}
\usepackage{color,cite,url}
\usepackage{textcomp, xcolor}

\newtheorem{assumption}{Assumption}

\def\BibTeX{{\rm B\kern-.05em{\sc i\kern-.025em b}\kern-.08em
    T\kern-.1667em\lower.7ex\hbox{E}\kern-.125emX}}
\begin{document}
\title{Policy Poisoning in Batch Learning for Linear Quadratic Control Systems via State Manipulation
\thanks{This work was supported in part by the National Science Foundation under Grant ECCS-2138956, and in part by a Faculty Research Grant from Fordham Office of Research.}
}

\author{\IEEEauthorblockN{Courtney M. King}
\IEEEauthorblockA{\textit{Computer \& Information Sciences Dept.} \\
\textit{Fordham University }\\
New York, NY 10023 \\
cking74@fordham.edu}
\and
\IEEEauthorblockN{Son Tung Do}
\IEEEauthorblockA{\textit{Computer \& Information Sciences Dept.} \\
\textit{Fordham University }\\
New York, NY 10023 \\
sdo@fordham.edu}
\and
\IEEEauthorblockN{Juntao Chen}
\IEEEauthorblockA{\textit{Computer \& Information Sciences Dept.} \\
\textit{Fordham University }\\
New York, NY 10023 \\
jchen504@fordham.edu}}
\maketitle
\begin{abstract} 
In this work, we study policy poisoning through state manipulation, also known as sensor spoofing, and focus specifically on the case of an agent forming a control policy through batch learning in a linear-quadratic (LQ) system. In this scenario, an attacker aims to trick the learner into implementing a targeted malicious policy by manipulating the batch data before the agent begins its learning process. An attack model is crafted to carry out the poisoning strategically, with the goal of modifying the batch data as little as possible to avoid detection by the learner. We establish an optimization framework to guide the design of such policy poisoning attacks. The presence of bi-linear constraints in the optimization problem requires the design of a computationally efficient algorithm to obtain a solution. Therefore, we develop an iterative scheme based on the Alternating Direction Method of Multipliers (ADMM) which is able to return solutions that are approximately optimal. Several case studies are used to demonstrate the effectiveness of the algorithm in carrying out the sensor-based attack on the batch-learning agent in LQ control systems. 
\end{abstract}


\section{Introduction}\label{sec:introduction}
Sensors have long been an invaluable tool to automation, forecasting, and efficiency \cite{borenstein1997mobile}, and they continue to grow in number and potential applicability. 
Our society is highly dependent on sensors to give frequent, consistent, and accurate information. Unfortunately, as AI-based systems become increasingly more popular and complex, so do their vulnerabilities. The sensor is a hot target as such systems rely on large data streams to operate, with sensors performing what is arguably the most important role in the data collection process. The threat of sensor-based attacks is severe \cite{davidson2016controlling}, and these types of attacks are especially prominent in situations where the sensors are deployed, and the data is sent to different control centers for subsequent processing.  

A control system can involve multiple components, and a system that uses learning-based approaches for decision-making, e.g., reinforcement learning, requires large amounts of data, having multiple data streams flowing into the system and between its components. This can leave the system vulnerable if the data becomes corrupted in storage or at any point during its transmission. An attack can effectively be carried out by manipulating the stored data before the agent even begins the learning process or at other stages if any channels within the system are left unprotected. For a typical multi-variable control system, it is even worse if the measurements of these variables (e.g., state and control) can be selectively altered to craft a strategic attack.  

To devise an effective learning-based mechanism on a control system, an agent requires information about the current state, action, and cost incurred at each point over a horizon. This data can be collected in batches, with these relevant measurements being taken from its sensors, actuators, and other components, and then used to obtain control policies to automate processes. Such a procedure is highly dependent on the quality of the data. Thus, it is extremely important that the measurements from the system are accurate and that the batch data is not corrupted. If a system is not strongly secured, this presents a weakness that could potentially be exploited by attackers who wish to damage or even inject their own controls into the system's automation. Sensor spoofing is one such case that attempts to manipulate the information collected from the system's sensors \cite{ zhang2017strategic}. 
In the case of learning a control policy from batch data, manipulating sensor data can result in control policies that are undesirable or even catastrophic. If an adversary has direct access to the batch data, not only can it significantly alter the intended goal of a learner, but also complete the attack strategically to avoid detection, allowing it to persist on the system \cite{liu2020secure}. In some cases, an attacker may even use reverse engineering to force a learner to uncover its own malicious policy instead \cite{rakhsha2020policy}. 

In this study, we focus on the case of sensor manipulation occurring within a linear quadratic (LQ) system during its batch-learning phase. The learner aims to devise an optimal control policy based on the sampled data, while the attacker aims to trick the learner into learning a malicious policy by poisoning the data samples. With perfect information of the system's learning process, the attacker alters the state values in the batch dataset such that the learner will learn the malicious policy instead of the optimal policy it expects. The attack is carried out strategically, meaning that the adversarial modifications are sparse and the state values are altered as minimally as possible for the attacker to achieve its goal while avoiding detection. This specific case of policy poisoning is formulated as an optimization problem, which poses additional challenges to solve, one of which being bi-linear constraints. We develop an algorithm that makes use of a heuristic method known as Alternating Direction Method of Multipliers (ADMM) to compute the solution and orchestrate the attack. A number of case studies are used to demonstrate the feasibility of the attack model and the effectiveness of the proposed algorithm.

\subsection{Related Works}
As the use of AI has become increasingly more common in ordinary applications, so have the threats posed by these techniques \cite{liu2022trustworthy}. An adversary can carefully devise various attacks such as advanced persistent threats \cite{chen2017security}, denial of services (DoS) \cite{yuan2015resilient}, false data injection (FDI) \cite{lin2021secure}, and sensor spoofing \cite{ zhang2017strategic}. Such attacks can occur in a variety of settings, manipulating critical systems on which we depend on a daily basis, from automated vehicles \cite{chen2019control} to medical and other IoT devices \cite{pawlick2017strategic}. Adversarial attacks on learning-enabled dynamic systems have also become ubiquitous. Understanding these attacks and developing countermeasures to guarantee the safety of the systems are critical \cite{brunke2022safe}. Previous works have explored policy poisoning attacks occurring via manipulation of the system's cost measurements in a discrete-time control system \cite{huang2022reinforcement, ma2019policy}. Our work focuses on the case of policy poisoning through sensor data manipulation on a batch-learning agent in the continuous-time setting, which we believe to be a more realistic and applicable attack scenario. 

\subsection{Organization of the Paper}
This paper is organized as follows. Section \ref{sec:preliminaries} defines the environment of the problem, introducing the LQ system and batch-learning process. Section \ref{sec:attack} develops the attack model and formulates the policy poisoning scheme as an optimization problem. Section \ref{sec:solutionmethod} presents the algorithm designed to obtain the optimal parameters to craft the attack on the batch learner. Section \ref{sec:casestudies} shows several case studies, demonstrating the effectiveness of the proposed attack model and implemented algorithm. Section \ref{sec:conclusion} concludes the study and gives suggestions for follow-up research. 

\subsection{Notations}
Let $\mathbb{R}_{+}$ denote the set of positive real numbers. Let $\mathbb{S}^{n}$ define the set of symmetric matrices of order $n$, and further let $\mathbb{S}^{n}_{+}$, $\mathbb{S}^{n}_{++}$ denote those which are also positive semi-definite or positive definite, respectively. The Frobenius norm of a matrix $M$ is denoted by $||M||_{F}$ (if the subscript is omitted, it denotes the Euclidean norm). The superscript $T$ denotes the transpose of a matrix. $I_n$ stands for the $n$-dimensional identity matrix. When the context is clear, we use $x, u, c$ instead of $x(t), u(t), c(t)$ to simplify notations. Let $M_{k}$ denote the value of $M$ after the $k$th iteration. Let ${M \succ 0}$ and ${M \succeq 0}$ denote that $M$ is positive definite and positive semi-definite, respectively. 

\section{Preliminaries}\label{sec:preliminaries}
This section defines the problem environment and surrounding context, provides essential background information, and introduces the batch-learning process.

\subsection{LQ Control Basics}
Consider a continuous-time linear dynamical system: 
\begin{equation}\label{eqn:system}
\begin{aligned}
    & \dot{x} = Ax+Bu,  \\
    & x(0) = x_0,
\end{aligned}
\end{equation}
where $x \in \mathbb{R}^{n}$, $u \in \mathbb{R}^{m}$ are state and control input vectors, respectively, and $A \in \mathbb{R}^{nxn}$ and $B \in \mathbb{R}^{nxm}$ are system matrices. 

The cost function to minimize is defined as
\begin{equation}\label{eqn:cost}
    J(x_{0},u)=\int_{0}^{\infty} (x^{T}Qx+ u^{T}Ru)dt,
\end{equation}
where $Q \in \mathbb{S}^{n}_{+}$, $R \in \mathbb{S}^{m}_{++}$ are the pre-defined cost matrices associated with the state and action, respectively. For convenience, we define the instantaneous cost in \eqref{eqn:cost} as
\begin{equation}\label{eqn:quad_loss}
\begin{aligned}
    & c(t) = x(t)^{T}Qx(t) + u(t)^{T}Ru(t).
\end{aligned}
\end{equation}
The optimal value function is defined as:
\begin{equation}\label{eqn:valuefunc}
\begin{split}
    V^{*}(x): &= \underset{u}{\text{min}} \ 
        \int_{0}^{\infty} (x^{T}Qx+ u^{T}Ru)dt\\
        &= x^TPx,
\end{split}
\end{equation}
where $P \in \mathbb{S}^{n}$.
We consider the feedback control policy that maps the current state to an action. Specifically, the control input is linear in the state:
\begin{equation}\label{eqn:control}
\begin{aligned}
    u=Kx,\\
\end{aligned}
\end{equation}
where $K\in\mathbb{R}^{m\times n}$.
We make the following assumption about the system.
\begin{assumption}\label{assm:stable}
The system described in \eqref{eqn:system} is stabilizable, i.e., the set
    \begin{equation}
        \mathcal{F} = \{ K | A+BK \ is \ stable \}
    \end{equation}
is non-empty.
\end{assumption}
Based on \eqref{eqn:valuefunc}, it is well known that the optimal feedback control policy admits the following form:
\begin{equation}\label{eqn:optcontrol}
\begin{aligned}
    & u^*= -R^{-1}B^{T}Px = Kx, \\
\end{aligned}
\end{equation}
where $P$ is obtained by solving the continuous algebraic Riccati equation (CARE):
\begin{equation}\label{eqn:ricatti}
    A^{T}P+PA-PBR^{-1}B^{T}P+Q=0.
\end{equation}

\subsection{Batch Learning}
In batch learning, samples are taken from an existing system to form a dataset of inputs and outputs. In each batch dataset $\mathcal{D}$, we assume that there are $N$ data points:
\begin{equation}\label{eqn:dataset}
\begin{aligned}
    & \mathcal{D} = \{(x_{k}, u_{k}, c_{k}) \textit{ for } k={0,1,2..N-1}\}, 
\end{aligned}
\end{equation} 
where 
\begin{equation*}
\begin{aligned}
    x_k \triangleq x(k\Delta t), \
    u_k \triangleq u(k\Delta t), \
     c_k \triangleq c(k\Delta t).
\end{aligned}
\end{equation*} 
Here, $\Delta t \in \mathbb{R_+}$ is the chosen sampling interval. Each data point is sampled from the system given in \eqref{eqn:system}. 

The goal of a learner is to first estimate the unknown system matrices from the batch dataset and then find the optimal feedback control \eqref{eqn:optcontrol}. 
We have the following assumption on the data-generation process.
\begin{assumption}\label{assm:zero_order}
    The dataset is generated under a zero-order hold, i.e., $u(t)$, for $t\in [k\Delta t,(k+1)\Delta t)$, $\forall k=0,1,..,N-1$, is held constant between each sampling instance.
\end{assumption} 

Additionally, it is required that the sampling time interval $\Delta t$ be chosen such that the spectral radius of $A$ is less than $\frac{1}{\Delta t}$. This constraint ensures successful learning of the underlying dynamic system based on data, as detailed in Section \ref{subsec:sys_id}.

\section{Policy Poisoning by State Manipulation}\label{sec:attack}
The learner uses the batch data to form its control policy, $K$, naturally assuming that the data is untouched and that its procedure will produce a desirable policy in its given environment. The attacker has full knowledge of the learner's processes to compute $K$ and plans to carry out the attack before the learner imports the batch data. The attacker's goal is to trick the learner into learning a malicious policy $K^{\dagger}$ that deviates from $K$, which could ultimately guide the system to taking unintended controls, based on \eqref{eqn:control}. Such an attack can be enabled by modifying the state measurements in the batch dataset. More specifically, the attacker can manipulate the control policy learning process by poisoning the benign dataset by replacing the state $x$ with deceptive state $x^{\dagger}$. 

\subsection{Learning the System Dynamics}\label{subsec:sys_id}
To devise an optimal attack, the attacker should follow the same procedures as the learner beforehand. As the learner is only given the batch dataset, he has to first perform identification of the underlying system dynamics. Thus, as an intermediate step, the attacker must perform the same system identification on the original dataset.
This is achieved through a two-step process that first estimates a discrete-time model from the samples and then determines an equivalent continuous-time model \cite{sinha2000identification}, summarized as follows.\\
\textbf{Step 1:} Learn a discrete-time model from samples.

The given samples are organized into system inputs and outputs assuming the form:
\begin{equation}
\begin{aligned}
    & x_{k+1} = Fx_{k} + Gu_{k},
\end{aligned}
\end{equation}
where 
\begin{equation}\label{eqn:FG2AB}
\begin{aligned}
     F=e^{A \Delta t}, \ G= \int_{0}^{\Delta t} e^{A \tau}d\tau B. \\ 
\end{aligned}
\end{equation}
The system parameters $F$ and $G$ can be obtained using the least-square estimates:
\begin{equation}\label{eqn:estFG}
    (\hat{F},\hat{G})= 
    \underset{F,G}{\text{argmin}}
    \sum_{k=0}^{N-1} ||(Fx_{k} + Gu_{k}) - x_{k+1} ||^{2}.
\end{equation}
The samples are separated into vectors as follows: 
\begin{equation}
\begin{aligned}
    X \coloneqq \begin{bmatrix}
        x_{1}^{T} \\ x_{2}^{T} \\ . \\  . \\ x_{N}^{T}
        \end{bmatrix},\
    Z \coloneqq \begin{bmatrix}
        z_{0}^{T} \\ z_{1}^{T} \\ . \\ . \\ z_{(N-1)}^{T}
        \end{bmatrix},\ 
\end{aligned}
\end{equation} 
where $X$ is a vector formed by concatenating the output sequence of the sampled data, and $z_{k} = [ x^{T}_{k}, u^{T}_{k}]^{T}. $
Under the assumption that $Z^{T}Z$ is invertible, we can directly obtain the estimations of $F$ and $G$ using the following: 
\begin{equation}
\begin{aligned}
    &  [ \hat{F} , \hat{G}]^{T} = (Z^{T}Z)^{-1}ZX.
\end{aligned}
\end{equation}
\textbf{Step 2:} Transform the discrete-time model into continuous-time model. 

We currently obtain the estimation of $F$ and $G$. 
Finding matrices $A,\ B$ based on \eqref{eqn:FG2AB} requires computing the natural logarithm of a square matrix. This can be done using the indirect method summarized in Algorithm \ref{alg:indirect}.\footnote{In Algorithm \ref{alg:indirect}, we choose $\epsilon=0.01$ to achieve reasonable accuracy and convergence time.}
\begin{algorithm}[!t]
    \caption{Indirect Method for Continuous Transformation}\label{alg:indirect}
    \begin{algorithmic}[1]
    \State Given initial parameters $n_{iter}$, $\epsilon= 0.01 $
    \State Determine $L=F-I$
    \State Set $P_{0}=I, M_{0}=I$, and $i=0$
    \Repeat 
        \State $M_{i+1}= \frac{-iLM_{i}}{i+1}$
        \State $P_{i+1}= P_{i}+M_{i+1}$
        \State $i=i+1$ 
    \Until {$i=n_{iter}$ or $||P_{i+1} - P_{i}||_{F} \leq \epsilon$ }
    \State Let $A=\frac{PL}{\Delta t}$, $B=\frac{PG}{\Delta t}$ 
    \State \Return $A,B$  
    \end{algorithmic}
\end{algorithm}

The cost matrices $Q$ and $R$ may also need to be estimated from the batch data. These can be obtained simply from sampled data $x$, $u$, and $c$ as follows\footnote{In our problem setup, only the attacker performs this step. The learner is assumed to hold the true estimates, a priori.}: 
\begin{equation}\label{eqn:estQR}
    (\hat{Q},\hat{R})= 
    \underset{Q,R}{\text{argmin}}
    \sum_{k=0}^{N-1} ||(x_{k}^{T}Qx_{k} + u_{k}^{T}Ru_{k}) - c_{k}||^{2}.
\end{equation}

\subsection{Policy Poisoning }\label{subsec:policy_poisoning}
To trick the learner into learning $K^{\dagger}$, the attacker plans to modify the batch dataset by replacing the original sensor/state measurements, $x$, with poisoned counterparts, $x^{\dagger}$. Additionally, the attacker aims to modify the dataset as minimally as possible to avoid detection.
Determining $x^{\dagger}$ is a nontrivial process. The attacker first estimates $\hat{A},\hat{B}$ using the methods described in subsection \ref{subsec:sys_id}, then estimates $\hat{Q},\hat{R}$ from the original batch data by \eqref{eqn:estQR}. With these learned parameters along with the targeted policy $K^{\dagger}$, the attacker devises $x^{\dagger}$ by solving the following optimization problem:  
\begin{equation}\label{eqn:attackstate}
\begin{aligned}
    \min_{x^{\dagger}, \tilde{A}, P} \ \ \ 
    & \sum_{k=0}^{N-1}||x^{\dagger}_{k} - x_{k}||^{2} \\
    \text{s.t.}\qquad 
    & \tilde{A}^{T}P + P(\tilde{A} + \hat{B}K^{\dagger})+\hat{Q}=0, \\
    & \hat{R}K^{\dagger} = -\hat{B}^{T}P, \\
    & P \succeq 0, \\
    & \dot{x}^{\dagger}= \tilde{A}x^{\dagger}+ \hat{B}u, \\
    & x^{\dagger}(0)=x_{0}.
\end{aligned}
\end{equation}
Based on \eqref{eqn:attackstate}, note that the attacker only changes the measurements of $x_k$ in the batch dataset while leaving the samples of control $u_k$ and cost $c_k$ untouched, for $k=0,1,...,N-1$. 

Since $c_{k}$ depends on $x_{k}$, modifying the values of $x_{k}$ without similarly modifying $c_{k}$ could result in inconsistencies. This problem can be addressed in two ways. i): It can be assumed that the learner already has perfect knowledge of $Q$ and $R$ and does not estimate $Q$ and $R$ again based on the batch dataset.  ii): The cost samples can be subsequently modified according to $ c^{\dagger}_{k}= x_{k}^{\dagger T} \hat{Q}x_{k}^{\dagger} +  u_{k}^{T}\hat{R}u_{k}$. For consistency, the problem \eqref{eqn:attackstate} should include this modification together with \eqref{eqn:estQR} into the constraint. 
In this work, we assume that the first scenario i) holds to facilitate the analysis. 

The problem \eqref{eqn:attackstate} is challenging to address. The first difficulty is due to the inclusion of the system dynamics in its constraints. The second is the bi-linearity of the first constraint in \eqref{eqn:attackstate}. To address the first challenge, we propose the following two steps. 
\\
\textbf{Step 1:} The attacker first finds an optimal $\tilde{A}$ that leads to $K^{\dagger}$ by solving the following optimization problem:
\begin{equation}\label{eqn:attackA}
\begin{aligned}
    \min_{\tilde{A}, P} \ \ \ \ \
    & || \tilde{A}-\hat{A} ||^{2}_{F}  \\
    \text{s.t.} \qquad 
    & \tilde{A}^{T}P + P(\tilde{A} + \hat{B}K^{\dagger})+\hat{Q}=0, \\
    & \hat{R}K^{\dagger} = -\hat{B}^{T}P, \\
    & P \succeq 0.
\end{aligned} 
\end{equation}
\textbf{Step 2:} The attacker then generates the poisoned data according to the following dynamics:  
\begin{equation}
\begin{aligned}
    & \dot{x}^{\dagger} = \tilde{A}x^{\dagger}+\hat{B}u, \\
    & x^{\dagger}(0)=x_{0}.
\end{aligned}
\end{equation}
Note that the poisoned states are sampled based on the same $\Delta t$ as used in the original batch data generation. 

The second challenge left to solve \eqref{eqn:attackA} is its non-convexity, due to the bi-linear terms, $\tilde{A}^{T}P$ and $P\tilde{A}$, in its constraints. 
Solving this requires us to develop an efficient computational scheme to obtain an approximately optimal solution, which is detailed in the ensuing section. 

\section{Solution Method}\label{sec:solutionmethod}
This section will discuss the method used to find a solution to \eqref{eqn:attackA}  and obtain the optimal attack parameter given the challenges discussed in Section \ref{subsec:policy_poisoning}. This is done by developing an algorithm based on the distributed optimization technique known as ADMM \cite{boyd2011distributed}. 

The algorithm, summarized in Algorithm \ref{alg:admm}, has three distinct steps. It can be interpreted as one alternatively updating the decision variables as necessary in order to resolve the inseparable terms, ${\tilde{A}}$ and ${P}$. 

First, the augmented Lagrangian to \eqref{eqn:attackA} is defined as
\begin{equation}\label{eqn:lagrangian}
\begin{aligned}
    \mathcal{L}_{\mu}(\tilde{A},P,Z)= ||\tilde{A}-\hat{A}||^{2}_{F}+
        Tr(Z^{T}W)+I_{P}+\frac{\mu}{2}||W||^{2}_{F},
\end{aligned}
\end{equation}    
with 
\begin{equation}
\begin{aligned}
    W = \begin{bmatrix}
        \tilde{A}^{T}P + P(\tilde{A} + \hat{B}K^{\dagger})+\hat{Q} \\
        \hat{R}K^{\dagger} + \hat{B}^{T}P \\  
        \end{bmatrix}, 
\end{aligned}
\end{equation}     
and 
\begin{equation}
\begin{aligned} 
    & I_{P} = \left\{ 
        \begin{array}{rcl}
            0 & \mbox{if} & P \succeq 0 \\ 
            +\infty & \mbox{o.w.} \\
        \end{array},
    \right.
\end{aligned}
\end{equation}
where ${Z}$ is a dual variable with an appropriate dimension matching $W$, and $\mu>0$ is a penalty parameter. Matrix $W$ captures the constraints in \eqref{eqn:attackA} and $I_{P}$ is the step function. 

\begin{algorithm}[!t]
    \caption{ADMM-based Approach}\label{alg:admm}
    \begin{algorithmic}[1]
    \State Given initial parameters:
    $\tilde{A}_{0}$, $P_{0}$, $Z_{0}$ penalty parameter ${\mu}$, number of iterations $n_{iter}$
    \For{ $i=0,1,...n_{iter-1}$ } 
        \State Let $\tilde{A}_{i+1} = 
        \arg\min_{\tilde{A}} \mathcal{L}_{\mu} (\tilde{A},P_{i},Z_{i}) $  
        \State Let $P_{i+1} = 
        \arg\min_{P} \mathcal{L}_{\mu} (\tilde{A}_{i+1},P,Z_i) $
        \State Let $Z_{i+1}= $ 
        \begin{equation*} 
            Z_{i}+\mu 
            \begin{bmatrix}
                \tilde{A}_{i+1}^{T}P_{i+1}+P_{i+1}
                (\tilde{A}_{i+1}+\hat{B}K^{\dagger})+\hat{Q} \\
                \hat{R}K^{\dagger}+\hat{B}^{T}P_{i+1} \\ 
            \end{bmatrix} 
        \end{equation*}
    \EndFor\\
    \Return $ \tilde{A}_{n_{iter}} $
    \end{algorithmic}
\end{algorithm}

The three major steps in Algorithm \ref{alg:admm} are presented below.

\textbf { 1) $\tilde{A}$ Step: } The first update step can be expressed as solutions to the following convex optimization problem: 
\begin{equation}
 \tilde{A}_{i+1} = 
        \arg\min_{\tilde{A}}\ \mathcal{L}_{\mu} (\tilde{A},P_{i},Z_{i}),
\end{equation}
and more specifically: 
\begin{equation}\label{eqn:Astep}
\begin{aligned}
    \min_{\tilde{A}}\ &|| \tilde{A}-\hat{A} ||^{2}_{F} \ + \\
    & \frac{\mu}{2}
    \left\Vert
        \begin{bmatrix}
            \tilde{A}^{T}P_{i} + P_{i}(\tilde{A} + \hat{B}K^{\dagger})+\hat{Q}+\frac{1}{\mu}Z^{1}_{i}\\ \\
            \hat{R}K^{\dagger} + \hat{B}^{T}P_{i}+\frac{1}{\mu}Z^{2}_{i}\\ \\ 
        \end{bmatrix}
    \right\Vert^{2}_{F}, \\
\end{aligned}
\end{equation}
with $Z_{i} = (Z^{1}_{i}, Z^{2}_{i}),$ where $Z_i$ comprises two components, $Z^{1}_{i}$ and $Z^{2}_{i}$. Notice that the trace term in the original Lagrangian is absorbed into the Frobenius norm term due to their equivalence. This step generates an $\tilde{A}$ that is close to $ \hat{A} $ while leading to the desired policy $K^{\dagger}$.

\textbf{ 2) $P$ Step: } The second step is to solve the following convex optimization problem: 
\begin{equation}
    P_{i+1} = 
        \arg\min_{P}\ \mathcal{L}_{\mu} (\tilde{A}_{i+1},P,Z_i),
\end{equation}
and more specifically: 
\begin{equation}\label{eqn:Pstep}
\begin{aligned}
    & \min_{P}\ 
    \left\Vert
        \begin{bmatrix}
            \tilde{A}^{T}_{i+1}P + P(\tilde{A}_{i+1} + \hat{B}K^{\dagger})+\hat{Q}+\frac{1}{\mu}Z^{1}_{i}\\ \\
            \hat{R}K^{\dagger}+\hat{B}^{T}P+\frac{1}{\mu}Z^{2}_{i}\\ 
        \end{bmatrix}
    \right\Vert^{2}_{F}, \\
    & \textrm{ s.t. } P \succeq 0.
\end{aligned}
\end{equation} 
This step results in $P_{i+1}$ which is an approximate solution to the Riccati equation, yielding the attacker's desired policy $K^{\dagger}$.

\textbf{3) $Z$ Step:} This step performs the dual update as follows:
\begin{equation}
\begin{aligned}
    Z_{i+1}&= Z_{i}+\\ 
    & \mu \begin{bmatrix}
        \tilde{A}^{T}_{i+1}P_{i+1} + P_{i+1}(\tilde{A}_{i+1} + \hat{B}K^{\dagger})+\hat{Q}+\frac{1}{\mu}Z^{1}_{i}\\ \\
        \hat{R}K^{\dagger}+\hat{B}^{T}P_{i+1}+\frac{1}{\mu}Z^{2}_{i}\\ \end{bmatrix}.
\end{aligned}
\end{equation}
The algorithm is guaranteed to converge to a stable point given that penalty parameter $\mu$ is sufficiently large \cite{gao2020admm}. Next, we will use several case studies to showcase the effectiveness of Algorithm \ref{alg:admm} for policy poisoning in batch learning.

\section{Case studies}\label{sec:casestudies}
In this section, we use several case studies to illustrate the
effectiveness of the developed policy poisoning scheme described in Section \ref{subsec:policy_poisoning} using the algorithm developed in Section \ref{sec:solutionmethod}. We leverage CVXPY \cite{diamond2016cvxpy} to solve the corresponding optimization problems when using Algorithm \ref{alg:admm}. 

\subsection{Case 1- Random Case}
We first demonstrate the attack on an LQ system with random parameters. The system starts at the initial state $x_0 = [0.5, -0.5, 0.5, -0.5]^T$. The sampling interval is $\Delta t = 0.01$, $Q=I_4,\ R = 0.5I_2$, and the system matrices are:
\begin{equation*}
A = 
\begin{bmatrix}
     0.59 &  0.13 & 0.33 & 0.76 \\
     0.63 & -0.33 & 0.32 & -0.05 \\
    -0.03 & 0.14 & 0.05 & 0.49 \\
    0.26 & 0.04 & 0.15 & 0.11
\end{bmatrix} 
, B = 
\begin{bmatrix}
     1.49 & -0.21 \\
     0.31 & -0.85 \\
    -2.55 & 0.65 \\
     0.86 & -0.74
\end{bmatrix}.
\end{equation*}
The optimal control policy can be obtained as:
\begin{equation*}
K^* = 
\begin{bmatrix}
    -2.87 & -0.38 & -0.34 & -1.24 \\
    4.32 & 1.14 & 3.63 & 4.71
\end{bmatrix}.
\end{equation*}
The learned system matrix $A$ based on the uncompromised dataset is:
\begin{equation*}
\hat{A} = 
\begin{bmatrix}
    0.60 &  0.13 & 0.33 & 0.76 \\
     0.63 & -0.33 & 0.32 & -0.05 \\
    -0.03 & 0.14 & 0.05 & 0.49 \\
    0.26 & 0.04 & 0.15 & 0.11
\end{bmatrix}.
\end{equation*}
With imperfect information on the system dynamics, the learner's expected control policy without attack is
\begin{equation*}
\hat{K} = 
\begin{bmatrix}
    -2.86 & -0.38 & -0.34 & -1.24 \\
    4.29 & 1.13 & 3.62 & 4.69
\end{bmatrix},
\end{equation*}
which closely resembles the optimal policy $K^{*}$.

\begin{figure}[!t]
  \centering
    \subfigure[States]{\includegraphics[width=0.49\columnwidth]{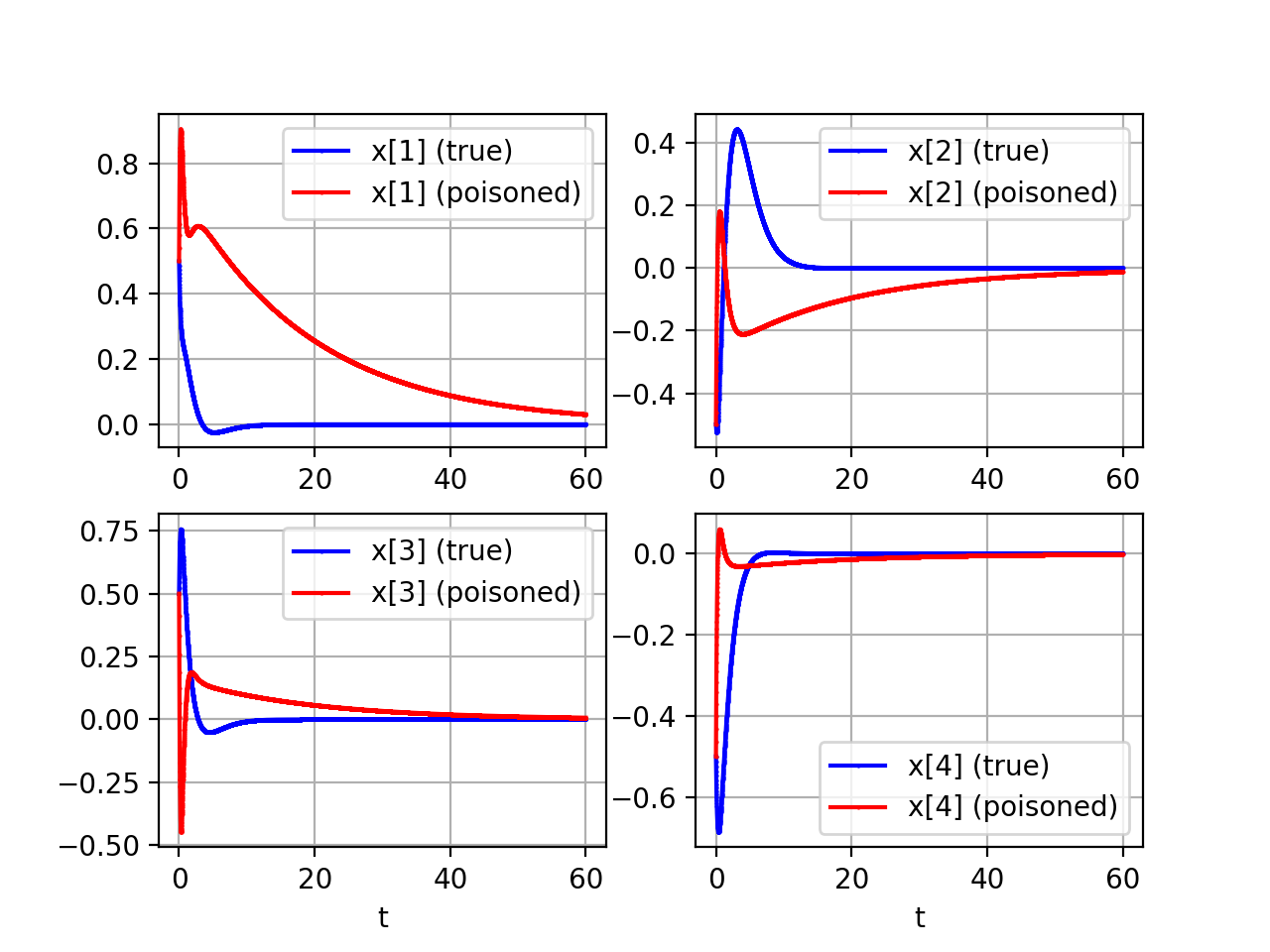}\label{fig:random_Figure_1}}
    \subfigure[Attack cost]{\includegraphics[width=0.49\columnwidth]{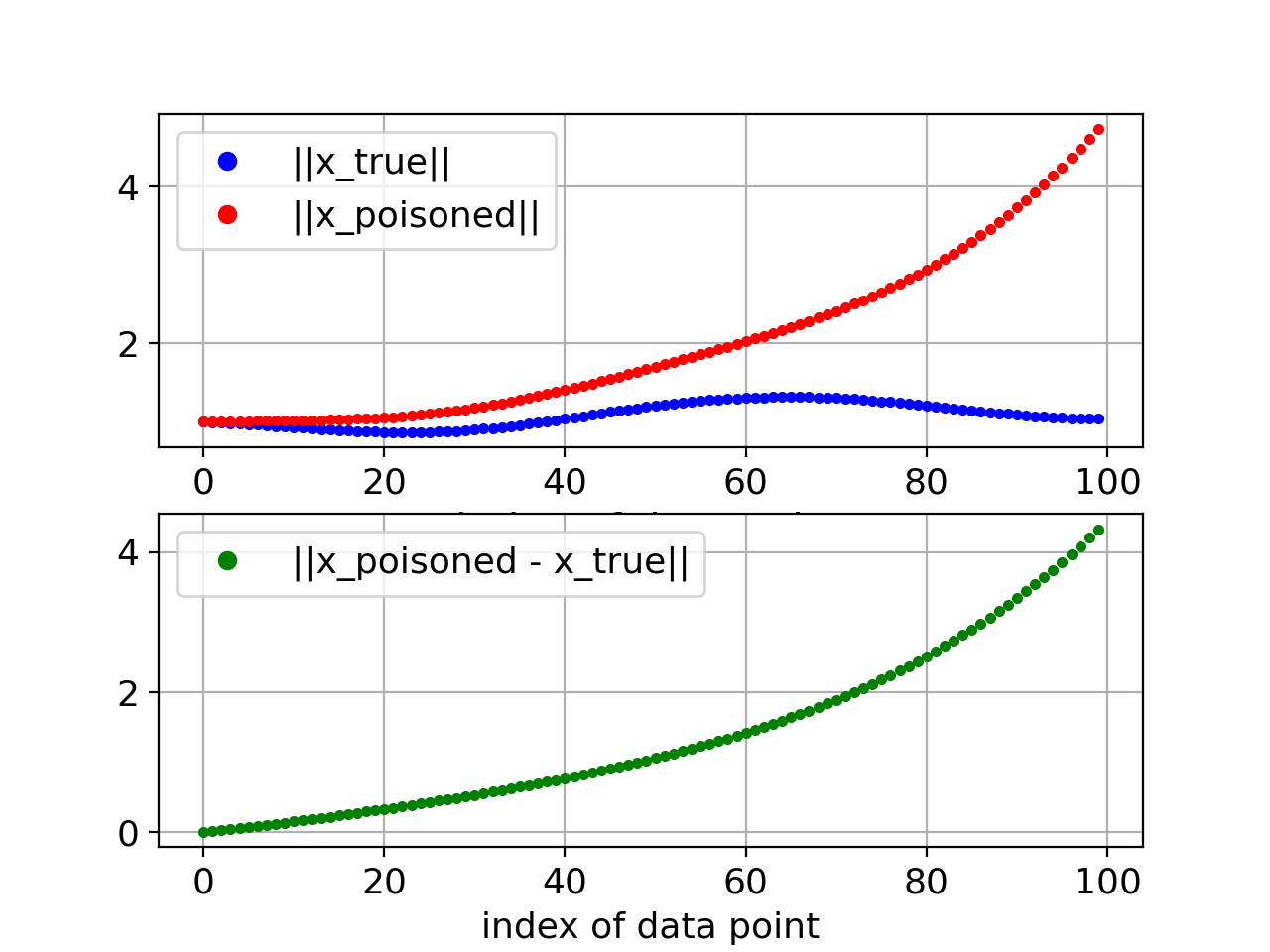}\label{fig:random_Figure_2}}
  \caption{Poisoning a random system}
  \label{fig:random}
\end{figure}

Assume that the attacker's target policy is
\begin{equation*}
K^\dag = 
\begin{bmatrix}
    -0.11 & 0.99 & 0.5 & -5.55 \\
    0.53 & 0.26 & 2.07 & 9.36
\end{bmatrix}.
\end{equation*}
The attacker then uses the steps in Section \ref{subsec:policy_poisoning} to generate the poisoned dataset. Based on Algorithm \ref{alg:admm}, it yields
\begin{equation*}
\Tilde{A} = 
\begin{bmatrix}
    0.17 & 0.05 & 0.02 & 0.41 \\
    0.41 & 0.24 & -0.29 & -0.35 \\
    -0.43 & 0.05 & 0.29 & 0.69 \\
    0.57 & 0.46 & 0.16 & 2.57
\end{bmatrix}.
\end{equation*}
The learner unknowingly learns from the poisoned dataset and reaches the policy:
\begin{equation*}
\hat{K} = 
\begin{bmatrix}
    -0.11 & 0.99 & 0.5 & -5.57 \\
    0.53 & 0.26 & 2.06 & 9.22
\end{bmatrix}.
\end{equation*}
Thus, the attacker successfully deceives the learner into learning the targeted policy. As shown in Fig. \ref{fig:random}, the attack causes a significant change in the system's state trajectories. In addition, the poisoned system takes much longer to stabilize.

\subsection{Case 2- Active Suspension System}

\begin{figure}[!t]
  \centering
  \subfigure[Position]{\includegraphics[width=0.49\columnwidth]{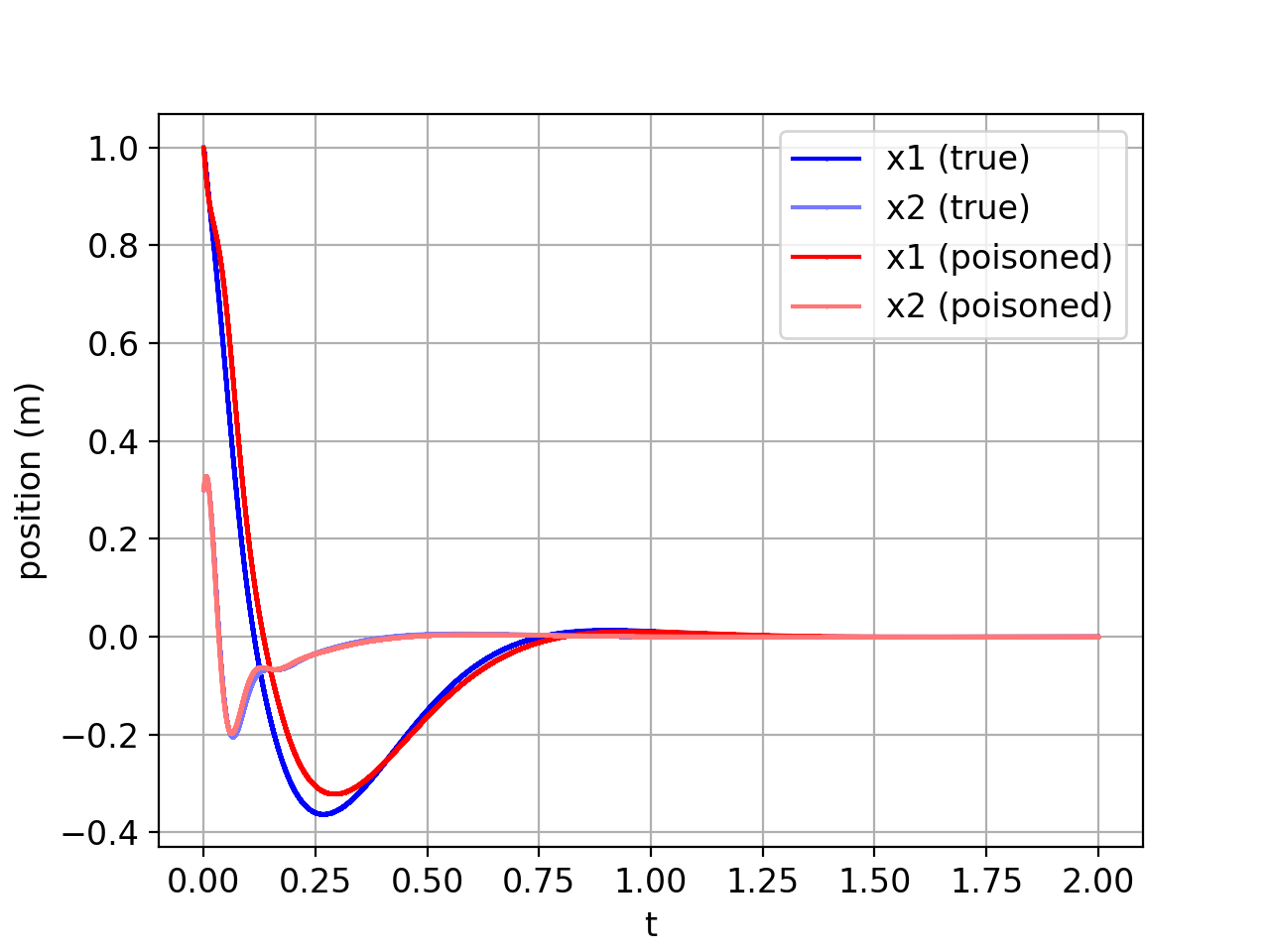}\label{fig:k_Figure_1}}
    \subfigure[Velocity]{\includegraphics[width=0.49\columnwidth]{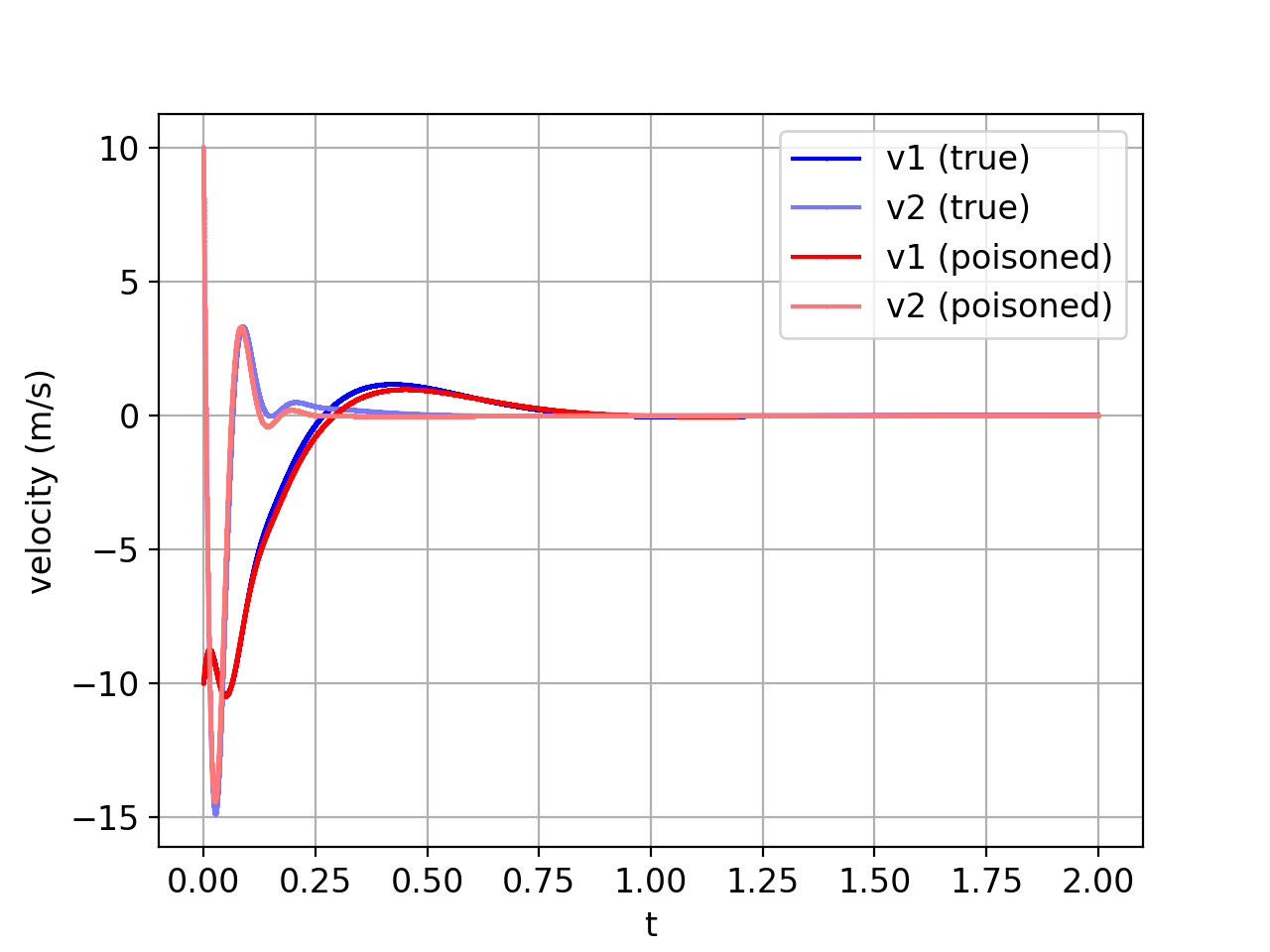}\label{fig:k_Figure_2}}
      \subfigure[Attack cost]{\includegraphics[width=0.7\columnwidth]{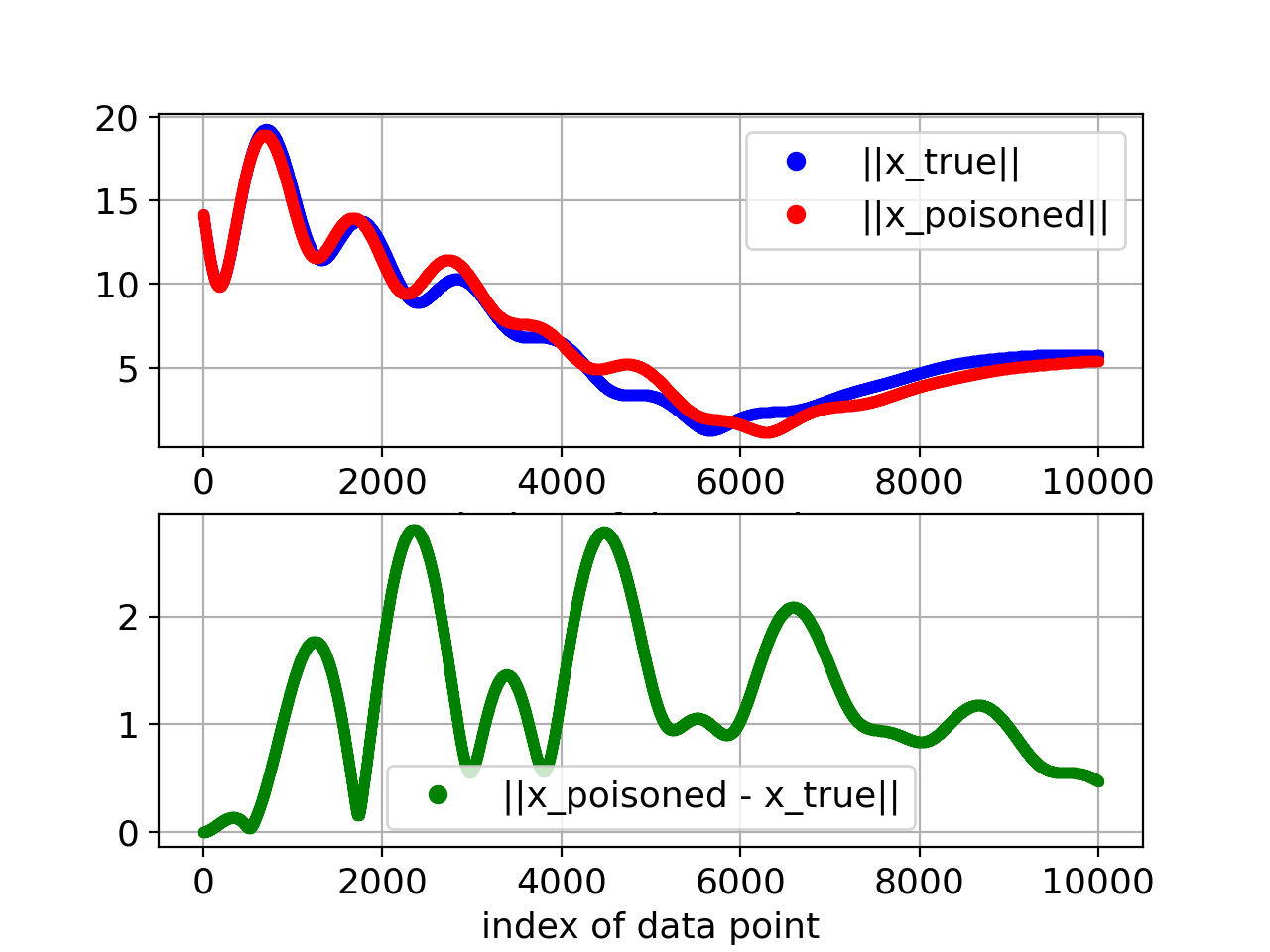}\label{fig:k_Figure_3}}
  \caption[]{Poisoning an active suspension system}
  \label{fig:suspension}
\end{figure}

Next, we attempt to poison the active suspension system on a vehicle. The active suspension system consists of a car body with mass $m_b$ and wheel assembly with mass $m_w$, connected by spring $k_s$ and damper $b_s$. Spring $k_t$ models the compressibility of the pneumatic tire. $x_b$ and $x_w$ are body travel and wheel travel, respectively. The force $f_s$ (in kilo Newtons) between the body and the wheel is the variable that can be controlled.
Denoting $(x_1, x_2, x_3, x_4) := (x_b, \dot{x}_b, x_w, \dot{x}_w)$, we get the linearized state-space representation \cite{ActiveSuspension}:
\begin{equation}
\begin{bmatrix}
    \dot{x}_1 \\
    \dot{x}_2 \\
    \dot{x}_3 \\
    \dot{x}_4
\end{bmatrix}
= 
\begin{bmatrix}
    0 & 1 & 0 & 0\\
    -\dfrac{k_s}{m_b} & -\dfrac{b_s}{m_b} & \dfrac{k_s}{m_b} & \dfrac{b_s}{m_b} \\
    0 & 0 & 0 & 1 \\
    \dfrac{k_s}{m_w} & \dfrac{b_s}{m_w} & \dfrac{-k_s-k_t}{m_w} & -\dfrac{b_s}{m_w}
\end{bmatrix}
\begin{bmatrix}
    x_1 \\
    x_2 \\
    x_3 \\
    x_4
\end{bmatrix}
+
\begin{bmatrix}
    0 \\
    \dfrac{10^3}{m_b} \\
    0 \\
    \dfrac{-10^3}{m_w}
\end{bmatrix}
\begin{bmatrix}
    f_s
\end{bmatrix},
\end{equation}
where the control is $u = f_s$. 
With  $m_b = 300\ kg,\ m_w = 60\ kg,\ b_s = 1000\ N/m/s,\ k_s = 16000\ N/m,\ k_t = 190000\ N/m$, the corresponding $A$ matrix is: 
\begin{equation*}
A = 
\begin{bmatrix}
    0 & 1 & 0 & 0\\
    -53.33 & -3.33 & 53.33 & 3.33 \\
    0 & 0 & 0 & 1 \\
    266.67 & 16.67 & -3433.33 & -16.67
\end{bmatrix}.
\end{equation*}
In addition, the initial state is $x_0 = [1, -10, 0.3, 10]^T$; the sampling interval is $\Delta t = 0.00005$; and the cost matrices are $Q = I_4 , R = 0.1I_1$. To this end, the optimal control policy is
\begin{equation*}
K^* = 
\begin{bmatrix}
    -0.31 & -2.57 & 30.6 & 2.22
\end{bmatrix}.
\end{equation*}
The estimated system matrix $A$ in batch learning admits
\begin{equation*}
\hat{A} = 
\begin{bmatrix}
    0 & 1 & 0 & 0\\
    -53.31 & -3.33 & 53.04 & 3.33 \\
    0 & 0 & 0 & 1 \\
    266.53 & 16.66 & -3431.87 & -16.74
\end{bmatrix}.
\end{equation*}
With imperfect information on the system dynamics, the learner's expected control policy is
\begin{equation*}
\hat{K} = 
\begin{bmatrix}
    -0.31 & -2.57 & 30.52 & 2.21
\end{bmatrix}.
\end{equation*}
The attacker aims to manipulate the learner into learning the parameter $A$ corresponding to different values of spring and damper coefficients. The attacker chooses $k_s = 2000\; N/m$, which yields
\begin{equation*}
K^\dag = 
\begin{bmatrix}
    -1.74 & -2.53 & 32.1 & 2.26
\end{bmatrix}.
\end{equation*}
The attacker then uses the steps in Section \ref{subsec:policy_poisoning} to generate the poisoned dataset.  Based on Algorithm \ref{alg:admm}, it yields
\begin{equation*}
\Tilde{A} = 
\begin{bmatrix}
    0 & 1.01 & 0 & 0\\
    -53.33 & -3.49 & 53.02 & 3.46 \\
    0 & 0 & 0 & 1.06 \\
    266.53 & 16.68 & -3431.87 & -16.73
\end{bmatrix}.
\end{equation*}
The learner ends up with the poisoned control policy:
\begin{equation*}
\hat{K} = 
\begin{bmatrix}
    -1.7 & -2.62 & 30.8 & 2.25
\end{bmatrix}.
\end{equation*}
Hence, the learner is successfully guided to the attacker's desired policy with a small amount of errors. From Fig. \ref{fig:suspension}, the poisoned system takes slightly longer to reach equilibrium. For a suspension system, this decreases the efficiency of the system and results in less comfort for the driver.

\section{Conclusion }\label{sec:conclusion}
In this study, we have pointed out a potential vulnerability of a batch-learning agent within an LQ system and developed a strategic attack model to effectively perform policy poisoning of the system. A computationally efficient algorithm has been designed to compute the optimal parameters to carry out the proposed attack on the system during its batch learning phase. Several case studies have demonstrated that the proposed policy poisoning scheme was successful in forcing the learner to learn a target policy desired by the attacker, significantly altering the system's performance. It is therefore crucial for an LQ system to actively protect its stored data, and specifically its sensor data, for trustworthy batch learning. Future work would include exploring methods of detection or protection from the proposed attack model and extending the poisoning scheme to online learning-based systems.
\bibliographystyle{IEEEtran}
\bibliography{IEEEabrv,references}
\end{document}